\documentclass{egpubl}
\usepackage{ICAT-EGVE2022}

 \WsPaperJoint         

\usepackage[T1]{fontenc}
\usepackage{dfadobe}  

\usepackage{cite}  
\BibtexOrBiblatex
\electronicVersion
\PrintedOrElectronic
\ifpdf \usepackage[pdftex]{graphicx} \pdfcompresslevel=9
\else \usepackage[dvips]{graphicx} \fi

\usepackage{egweblnk} 
\usepackage{cite}
\usepackage{CJKutf8}
\usepackage{here}
\usepackage{url}
\usepackage[hang,small,bf]{caption}
\usepackage[subrefformat=parens]{subcaption}
\title{AR Object Layout Method Using Miniature Room\\ Generated from Depth Data}
\author[K. Ihara and I. Kawaguchi]
{\parbox{\textwidth}{\centering K. Ihara$^{1}$ and I. Kawaguchi$^{1}$}
\\
{\parbox{\textwidth}{\centering $^1$University of Tsukuba, Japan}}
}
\begin{document}
\begin{CJK}{UTF8}{min}
\maketitle
\begin{abstract}
In augmented reality (AR), users can place virtual objects anywhere in a real-world room, called AR layout.
Although several object manipulation techniques have been proposed in AR, it is difficult to use them for AR layout owing to the difficulty in freely changing the position and size of virtual objects.
In this study, we make the World-in-Miniature (WIM) technique available in AR to support AR layout.
The WIM technique is a manipulation technique that uses miniatures, which has been proposed as a manipulation technique for virtual reality (VR).
Our system uses the AR device's depth sensors to acquire a mesh of the room in real-time to create and update a miniature of a room in real-time.
In our system, users can use miniature objects to move virtual objects to arbitrary positions and scale them to arbitrary sizes.
In addition, because the miniature object can be manipulated instead of the real-scale object, we assumed that our system will shorten the placement time and reduce the workload of the user.
In our previous study, we created a prototype and investigated the properties of manipulating miniature objects in AR.
In this study, we conducted an experiment to evaluate how our system can support AR layout.
To conduct a task close to the actual use, we used various objects and made the participants design an AR layout of their own will.
The results showed that our system significantly reduced workload in physical and temporal demand.
Although, there was no significant difference in the total manipulation time.

\end{abstract}

\section{Introduction}
Currently, HoloLens2, MagicLeap, NrealLight, and other augmented reality (AR) devices are available, and are becoming increasingly popular.
In AR, users place virtual objects anywhere in a real-world room (henceforth "AR layout").
In AR layout, possible situations include a user arranging windows on a desk or sticking notes on a refrigerator.
However, with the existing user interface, AR layout is time-consuming and burdensome for users.
Therefore, several studies have been conducted to shorten the manipulation time and reduce the workload on users when placing virtual objects in a room. 

Several techniques have been proposed for manipulating virtual objects in AR, such as feature point-based techniques~\cite{ProjectiveWindows, SnapToReality}, as well as those employing room deformation~\cite{wall-based}.
The techniques that uses feature points have the disadvantage that virtual objects cannot be placed in areas where no feature points exist, such as next to or behind a certain virtual object.
The technique that use room deformation is difficult to manipulate with an accurate understanding of the original layout.

Therefore, in this study, we focused on the World-in-Miniature (WIM) technique~\cite{wim}, a virtual object placement technique in virtual reality (VR).
The WIM technique is a user interface technology that uses miniature models of virtual environments as maps and interaction spaces.
This technique allows users to manipulate virtual objects by manipulating miniature objects, and can help users manipulate virtual objects.
In addition, because the WIM technique can use two viewpoints, either the first-person viewpoint or the bird's eye viewpoint, it can assist in recognizing the position of the virtual object.
Because of these characteristics, the WIM technique has been used in navigation~\cite{WIM-navigation}, IoT manipulation~\cite{WIM-IoT}, living or working spaces design~\cite{WIM-vrroom, tangibleAR}, and annotating real objects in AR~\cite{annotateWIM}.
However, it has not been used for AR layout.
In this study, we employed the WIM technique in AR by creating a miniature of a room with depth sensors attached to the AR device.

In our previous study~\cite{IharaMini}, we created a prototype and conducted a preliminary experiment to obtain design guidelines for the system.
In our preliminary experiment, we did not evaluate the miniature room but the manipulability of the miniature objects in AR compared to hand-ray manipulation.
The results revealed the characteristics of manipulating miniature objects.
The use of miniature objects was effective for checking virtual objects' placement and moving manipulations. However, there was no significant difference in scaling manipulations.

In this study, we improved our system to solve the manipulation issues shown in our previous study.
In addition, we refined our system by improving the appearance of the miniature.

Thereafter, to investigate our system's usefulness in AR layout, we conducted an experiment under an environment that simulated the actual usage.
In this experiment, contrary to the previous manipulability study, participants performed the task of freely arranging multiple virtual objects according to a scenario that simulates actual usage. 
The results showed that our system significantly reduced workload in physical and temporal demand.
Although, there was no significant difference in the total manipulation time.

\section{Related work}
In this section, we first introduce classical virtual object placement techniques including the WIM technique.
Subsequently, we introduce research that use depth sensor to capture the room because we use depth sensor for creating a miniature of a room.
Finally, we introduce virtual object placement techniques in AR, and explain the challenges of these techniques and the purpose of our work.

\subsection{Classical virtual object placement techniques}
The most common technique for virtual object placement in VR is the ray-casting technique\cite{raycast}, emits virtual rays of light from the hand and uses collisions between the rays and objects to select and move virtual objects.
Pourpyrev et al.'s Go-Go manipulation technique~\cite{go-go} uses a nonlinear mapping of actual arm extension to a virtual arm. This allows the selection and manipulation of objects at distances that are out of reach with the actual arm.
Pierce et al.'s image plane technique~\cite{ImagePlane} maps virtual objects to an image plane. By selecting the virtual objects in the image plane, users can select virtual objects.
Stoakley et al.'s WIM technique~\cite{wim} copies a user's virtual room and shrinks it to make a miniature model of the room. By manipulating miniature objects in this miniature model, users can manipulate virtual objects in the room.

Some of these techniques are used in object manipulation in AR. For example, the ray-casting technique is used as a default manipulation in HoloLens2~\cite{hololens-handray}. The Go-Go manipulation technique is used for manipulating real objects~\cite{go-go-real, go-go-sar}. The image plane technique is used for virtual object placement in AR~\cite{ProjectiveWindows}. On the other hand, the WIM technique is used for navigation\cite{WIM-navigation}, IoT manipulation~\cite{WIM-IoT}, living or working spaces design~\cite{WIM-vrroom, tangibleAR}, and annotating real objects in AR~\cite{annotateWIM}, although it has not been used in virtual object placement in AR.

Therefore, in our work, we employed the WIM technique for virtual object placement in AR.

\subsection{Using depth data of the room}
To use the WIM technique in AR for virtual object placement, different from VR where it is easy to make a copy of the virtual environment, we need the 3D data of the real environment to make a miniature model of the room.
There is much research using depth sensors for capturing the 3D data of the room to support AR interaction.
For example, in Remixed Reality~\cite{RemixedReality}, they used multiple depth sensors to capture the room in real-time, and 
make the environment as changeable as VR.
In SceneCtrl~\cite{SceneCtrl}, they used the depth sensor attached to the AR device to make it possible to edit the real environment.
In RoomAlive~\cite{RoomAlive}, they used depth sensors and projectors to make the real room into an interactable AR space.
In this paper, we use the depth sensor attached to the AR device to create a miniature model of a room.

\subsection{Virtual object placement techniques in AR}
There are two techniques for placing virtual objects in AR: automatic placement and manual placement. 
Automatic placement techniques include automated methods based on geometry~\cite{flare} and automated techniques based on the meaning of the real objects placed in the room~\cite{SemanticAdapt}.
The automatic placement technique does not always place the items where the user intends them to be placed. Therefore, it is necessary to choose whether to use the automatic or manual placement technique depending on the case.

Manual placement techniques use room feature points.
In Lee et al.'s~\cite{ProjectiveWindows} technique, the user selects a distant object using the image plane technique, and when the virtual object is released, it is attached to a real-world wall while retaining its apparent size. Nuernberger et al.'s~\cite{SnapToReality} technique acquires the surfaces and edges of a real object and uses them as constraints to assist the user in placing the object.
However, placement cannot be made at an arbitrary position, such as next to a certain virtual object, because it cannot be made at a location without feature points.
In Chae et al.'s technique\cite{wall-based}, a virtual wall is superimposed on a real-world wall, and the user moves the wall back and forth to move the virtual objects in the room back and forth.
This assisted the user's perception of the depth direction and assisted in the placement of the virtual objects.
It is difficult to employ this technique for scaling manipulation. 
Therefore, it cannot be employed in AR layouts, where scaling manipulation is mandatory.
In our work, virtual objects can be placed anywhere in the room, with or without feature points, and can be scaled to any size.

\section{Contribution of this study}
In our study, we create a system that enables the use of the WIM technique in AR by creating a miniature of a room using depth sensors.
Contrary to previous studies, our system allows users to move virtual objects anywhere in the room, with or without feature points, and freely adjust these sizes.
Therefore, our system can be used in AR layout.
Furthermore, we consider that our system can support AR layout in shortening manipulation time and reducing workload owing to the characteristics of the WIM technique.

In our previous study~\cite{IharaMini}, we created a prototype and conducted preliminary experiments to investigate the properties of miniature object manipulation.
Preliminary experiments revealed that using miniature objects is effective for checking virtual objects' placement and moving manipulations.
However, the use of miniature objects was ineffective in scaling manipulation.
Therefore, in this study, we aim to improve the prototype to support scaling manipulation.
In addition, this preliminary experiment had the following limitation. 
As we aimed to investigate only the manipulability of miniature objects, we only used spheres in this experiment.
Although in the actual AR layout, users place various virtual objects, including 2D windows, 3D objects, and 3D icons.
In addition, to investigate only the manipulability, we predetermined the target position and size in the experiment.
Although in actual usage, users choose where they want to place the objects and choose the size of the objects of their own free will.
Because of the above points, the situation used in the preliminary experiment was not similar to actual usage situations.
Therefore, in this study, we conduct an experiment close to the actual use case in AR layout using various objects and making the participants design the layout of their own will.

In this study, the following contributions were made:
\begin{itemize}
    \item We created a system that enables the use of the WIM technique in AR.
    \item To investigate the usefulness of this system in AR layout, we conducted an experiment under similar experimental conditions to those in which the system will be used.
\end{itemize}

\section{System}
\subsection{System overview}
\begin{figure}[H]

\begin{center}
\includegraphics[width=0.45\textwidth]{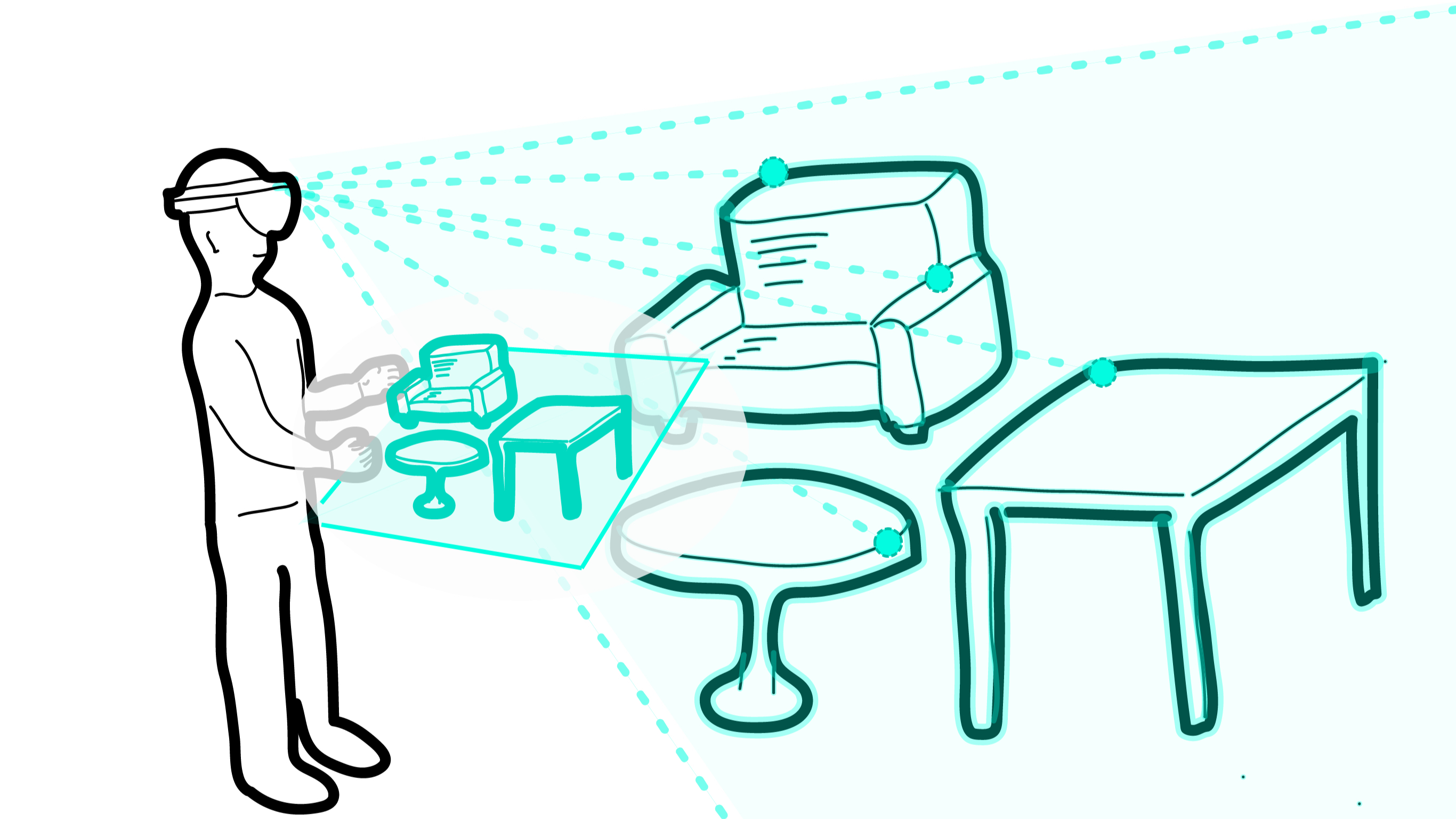} 
\caption{\label{overview}Overview of our system.}
\end{center}
\end{figure}

The miniature in our system consists of a miniature room and miniature objects.
A miniature room is a miniature of a user's real environment, which includes walls, floors, and real objects.
Our system creates a miniature room using the room's mesh generated using depth data (Fig.\ref{overview}).
Miniature objects are miniatures of virtual objects.
In our system, we show a miniature object in the miniature room when the user shows a virtual object in the room.
Our system moves miniature objects in conjunction with the virtual objects in the room.
Therefore, users can check the position of virtual objects in the room by checking the miniature objects.
In addition, users can change the position and size of virtual objects in the room by changing those of miniature objects.

\subsection{Implementation}
\begin{figure}[H]
\begin{center}
\includegraphics[width=0.4\textwidth]{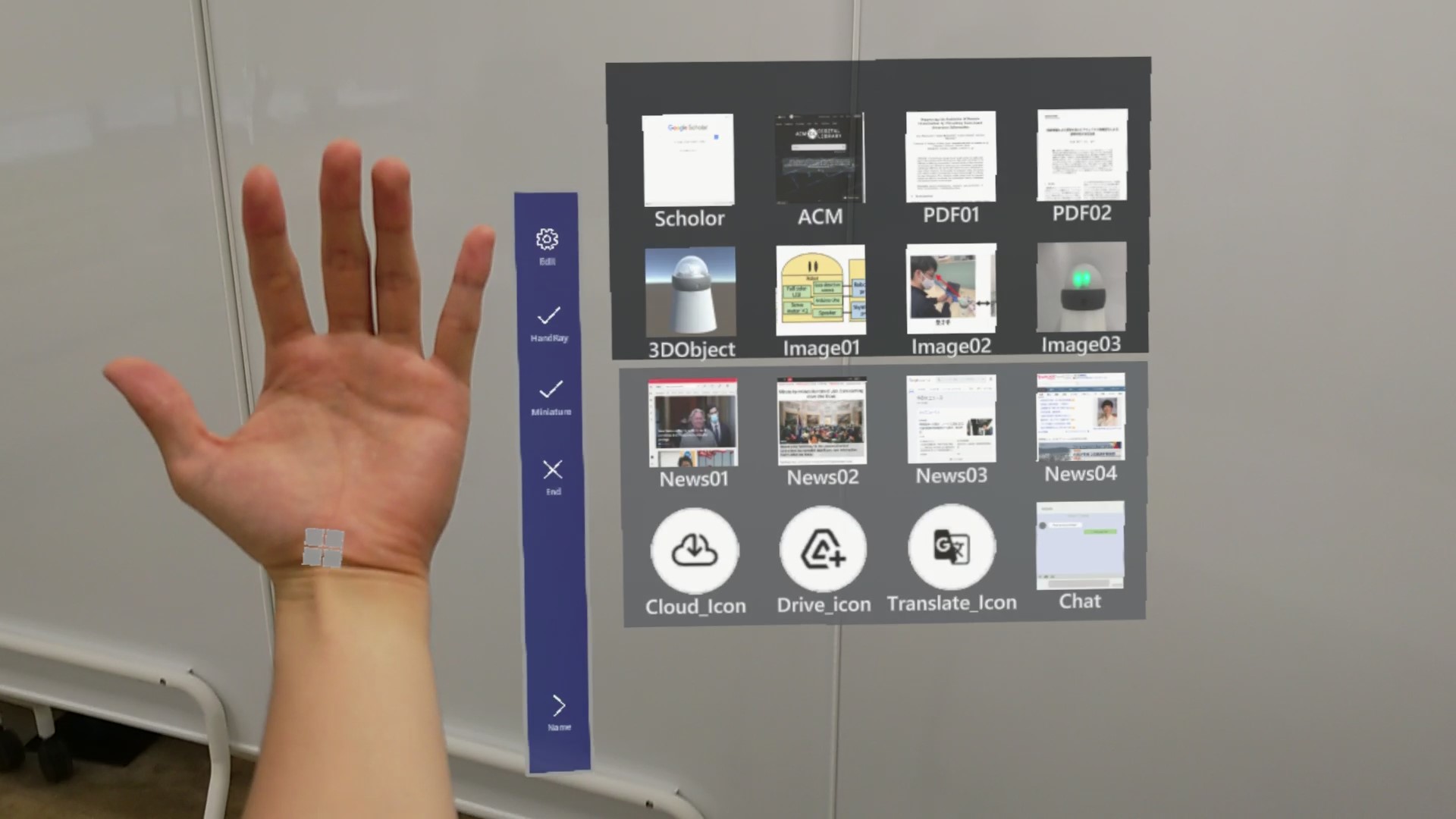} 
\caption{\label{handmenu}Hand menu.}
\end{center}
\end{figure}

We developed the system on Unity (version 2019.4.22f1) using Mixed Reality Toolkit (version 2.6.2, henceforth MRTK) and deployed it on HoloLens2.
Our system generates the room mesh using the Spatial Awareness System of MRTK, which generates a mesh of the room from depth data and updates it in real-time.
In addition, our system uses a hand menu that shows two UIs (Fig.\ref{handmenu}).
One of the UI shows the button to switch the manipulation mode between the miniature objects manipulation and miniature room manipulation.
The other UI shows the images of virtual objects that the users can place.
The corresponding object appears one meter in front of the user by pressing the image. 

\subsection{Generate the miniature room and objects}
\begin{figure}[H]

\begin{center}
\includegraphics[width=0.4\textwidth]{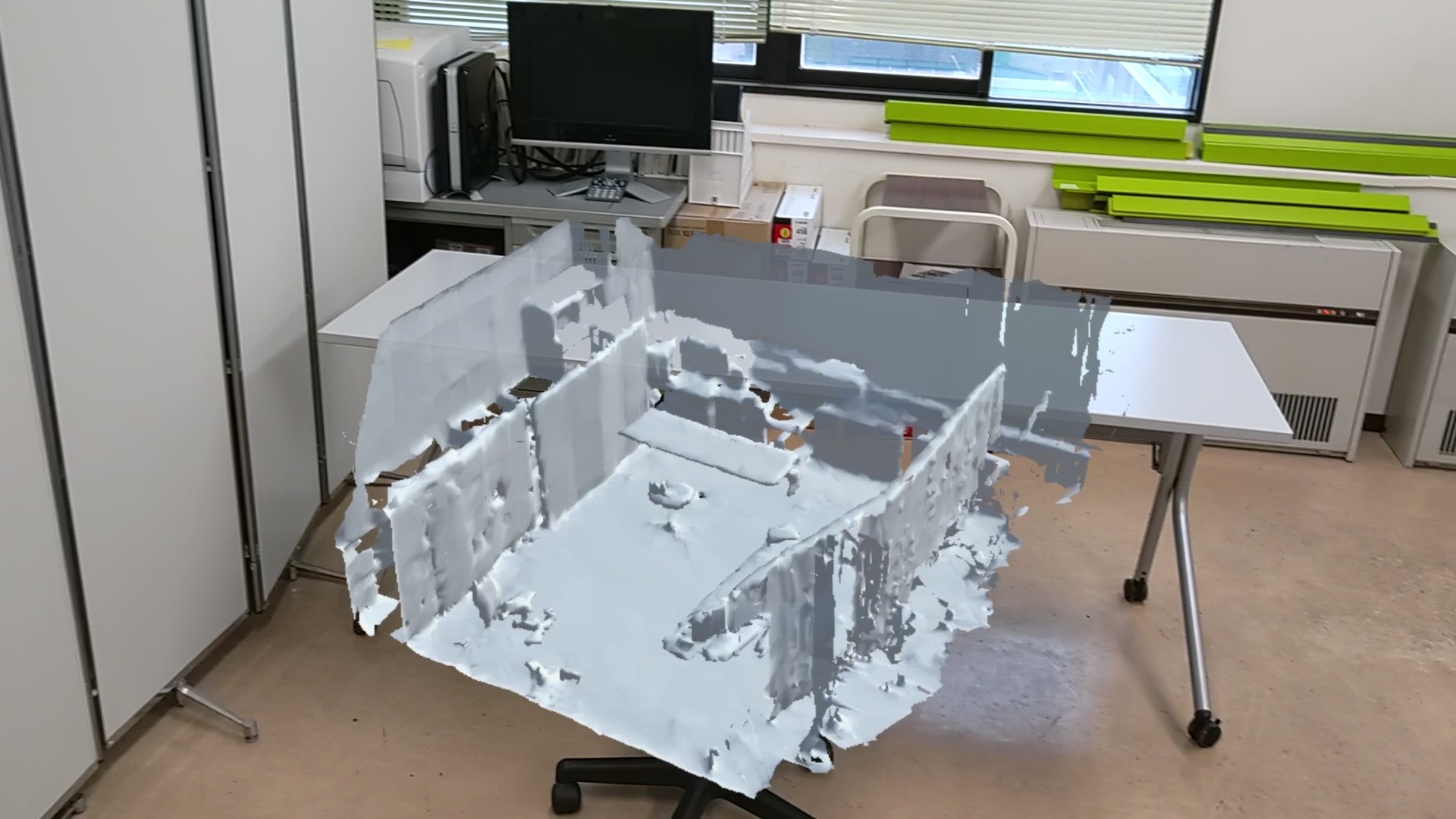} 
\caption{\label{miniature_room}Appearance of the miniature room.}
\end{center}
\end{figure}

We demonstrate the generation of the miniature room.
Our system shrinks the room mesh provided by the function of MRTK, called the Spatial Awareness System, to create the miniature room. 
This function creates the mesh of the room by using the depth data acquired from the depth sensor attached to the AR device.
In addition, this function is used fundamentally in most AR applications.
Therefore, the processing load using our system in AR application is limited.
Furthermore, because this function works in real-time, the mesh of the miniature is updated in real-time.
Although the system provides wireframe mesh, our system adds surfaces to aid users recognize the room (Fig.\ref{miniature_room}).

Our system generates a miniature object when the user puts a virtual object in the room.
When the user changes the position and size of the objects, those of the corresponding miniature objects also change.

\subsection{Manipulate the miniature room and objects}
\begin{figure}[H]

\begin{center}
\includegraphics[width=0.4\textwidth]{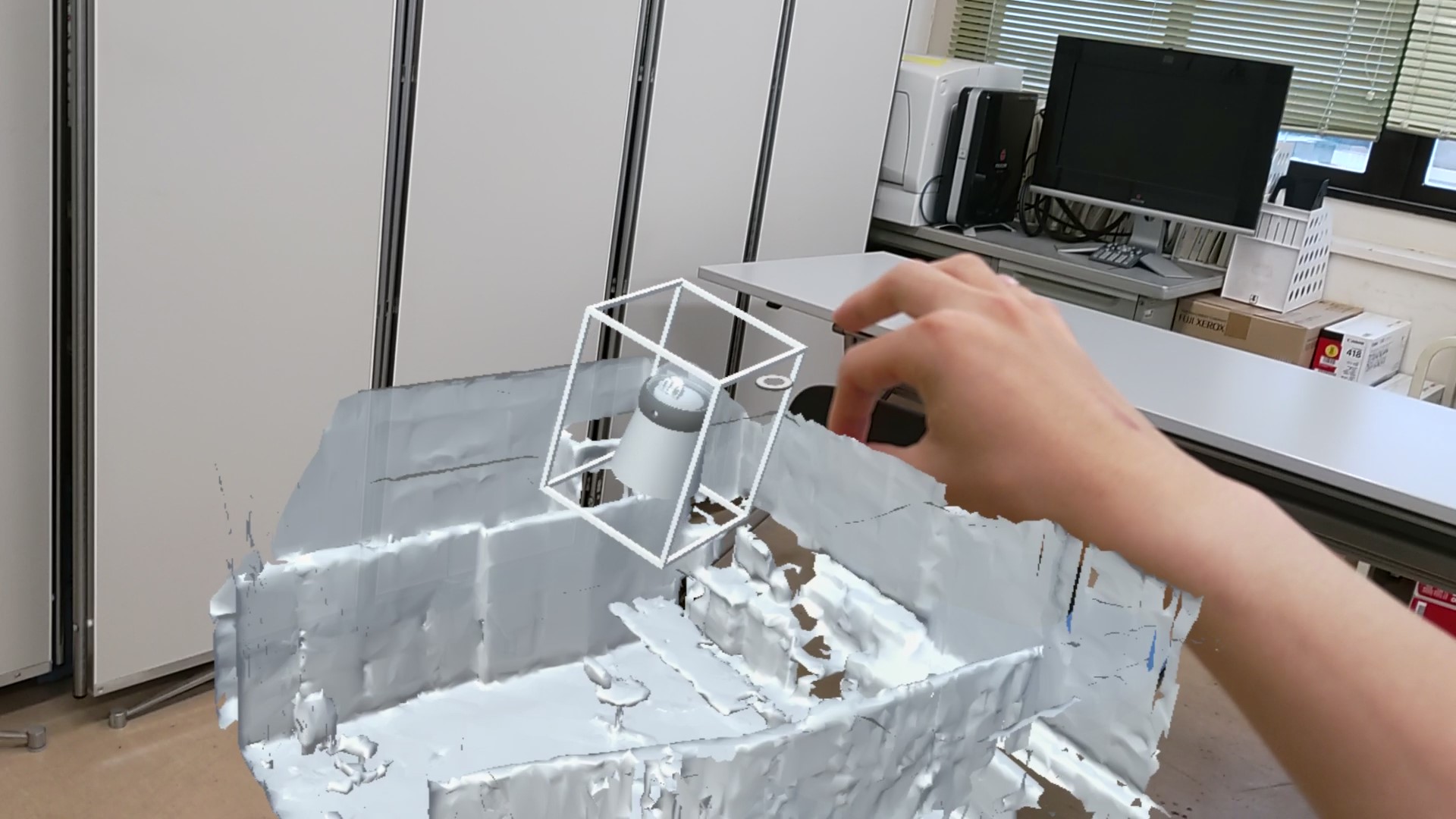} 
\caption{\label{boundingbox}Moving the miniature objects.}
\end{center}
\end{figure}

\begin{figure}[H]
\begin{center}
\begin{minipage}[b]{0.45\linewidth}
\includegraphics[keepaspectratio,scale=0.056]{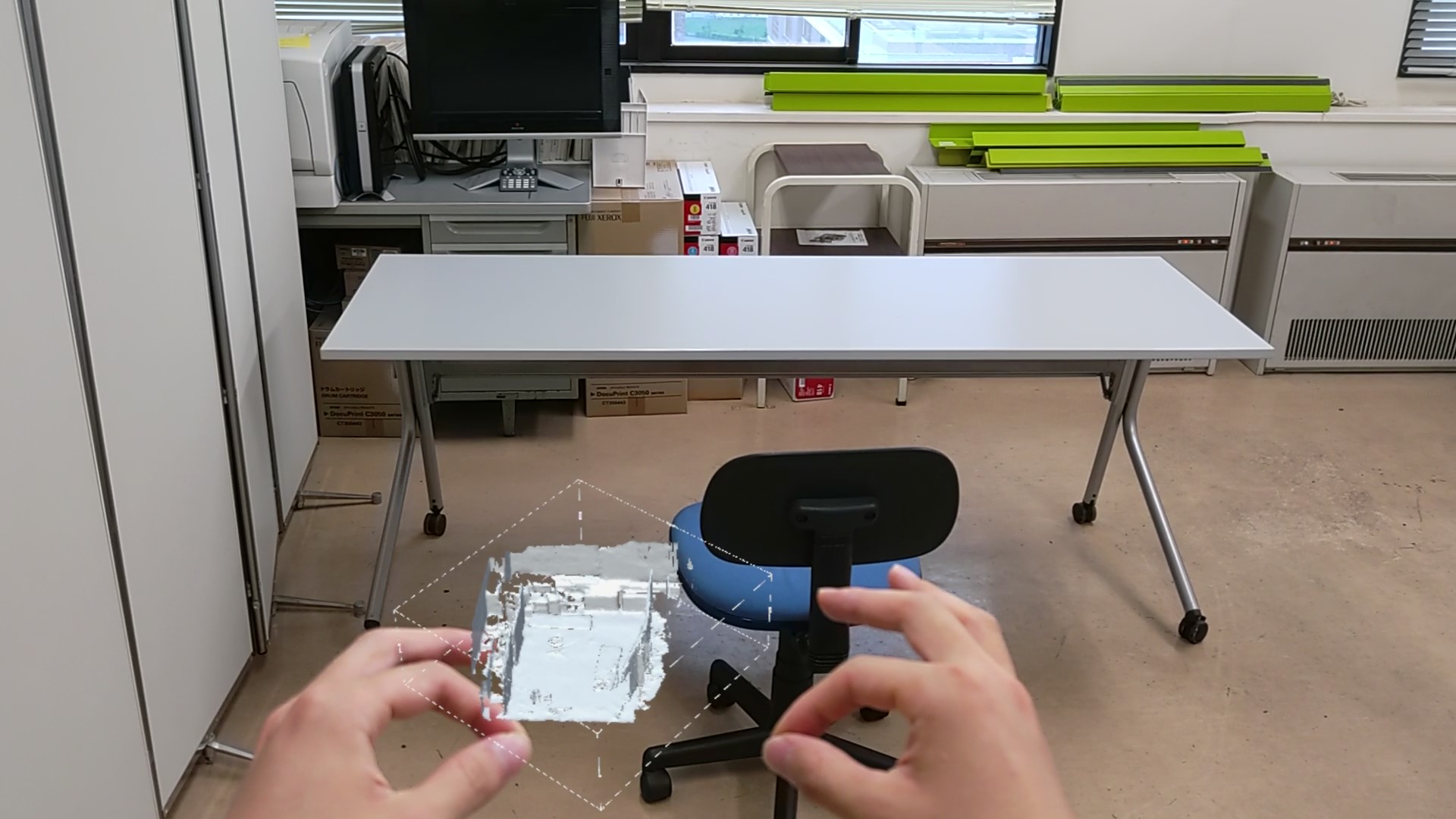} 
\end{minipage}
\begin{minipage}[b]{0.45\linewidth}
\includegraphics[keepaspectratio,scale=0.056]{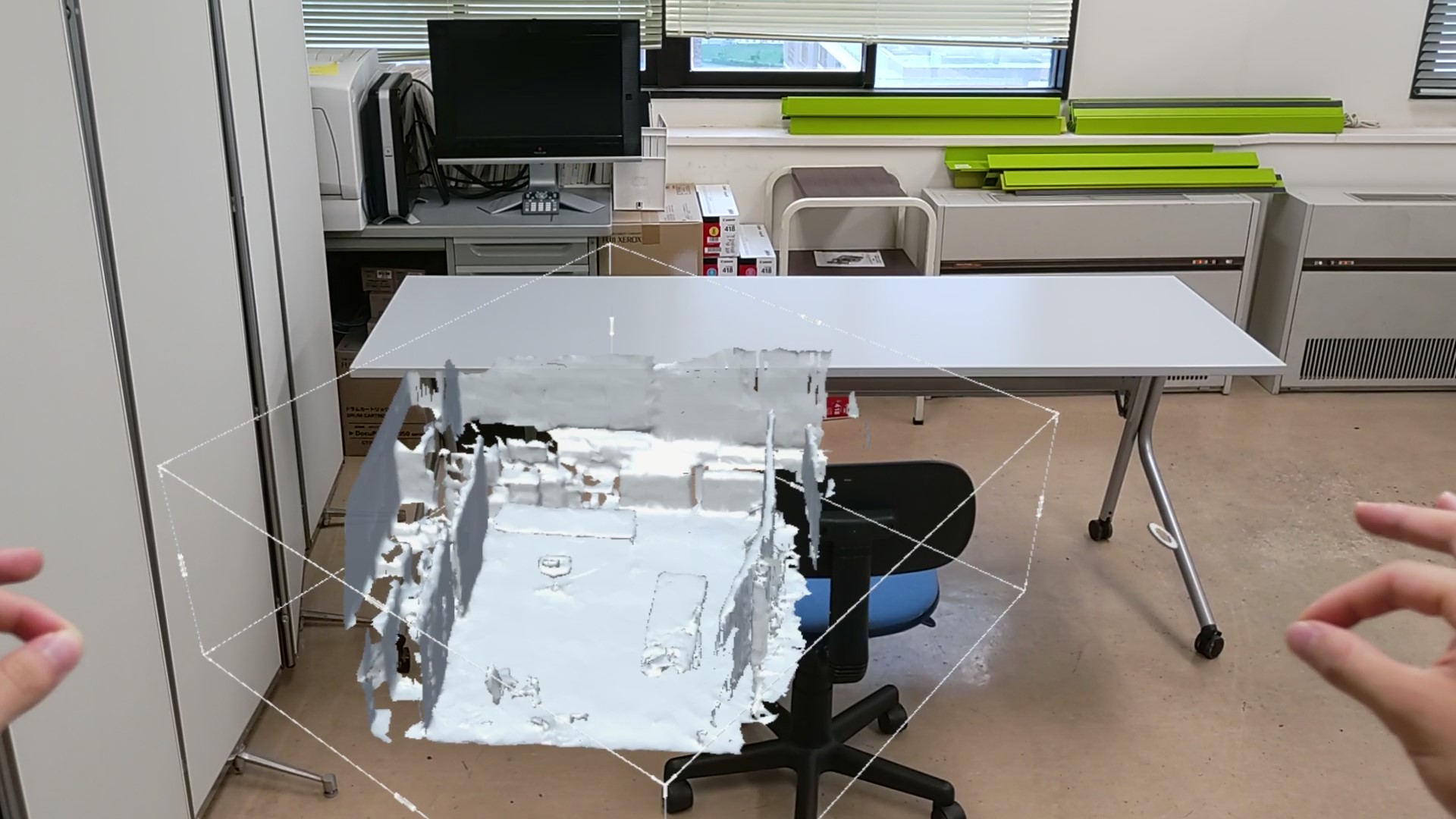} 
\end{minipage}
\caption{\label{scaling}Scaling the miniature room (left: before, right: after).}
\end{center}
\end{figure}

In our system, users can manipulate the miniature room and the miniature objects using the same manipulation procedures.
The miniature room and miniature objects can be moved by pinching them directly with one hand and moving the hand to a target position (Fig.\ref{boundingbox}).
Furthermore, they can be scaled by pinching them directly with two hands and changing the distance between two hands (Fig.\ref{scaling}).
Because the preliminary study showed that scaling manipulation is difficult, our system uses a bounding box that covers each the miniature room and miniature objects.

If a user tries to manipulate a miniature object while the bounding box is set on the miniature room, our system will recognize that the user is trying to manipulate the miniature room.
Therefore, our system requires users to switch between miniature room manipulation and miniature object manipulation.
To enable users to switch the manipulation mode, our system shows a virtual button in the hand menu.
When the user switches to the miniature room manipulation mode, a bounding box appear on the miniature room.
Conversely, when the user switches to the miniature objects manipulation mode, the bounding box on the miniature room disappears.

\section{Evaluation}
We evaluated how our system can support AR layout.
In AR layout, users place various virtual objects in a real environment and decide the position and size of the virtual objects as per their will.
Therefore, to conduct a task close to the actual use, we first used various virtual objects, including 3D objects, 2D windows, and 3D icons in the task.
Second, we did not specify the target position and size of the virtual objects and made the participants decide the position and size as per their will.

We made an application for this experiment on Unity2019.4.22f1 using MRTK2.6.2 and deployed it on HoloLens2.
We used a room that width, depth, and height are 3.2, 4.8, and 2.5 m, respectively.

\subsection{Conditions}
We compared the following two conditions.
\begin{description}
\item[Hand-Ray] Participants use only Hand-Ray manipulation 
\item[Miniature] Participants use only Miniature manipulation
\end{description}

\begin{figure}[H]
\begin{center}
\begin{minipage}[b]{0.49\linewidth}
\includegraphics[keepaspectratio,scale=0.0605]{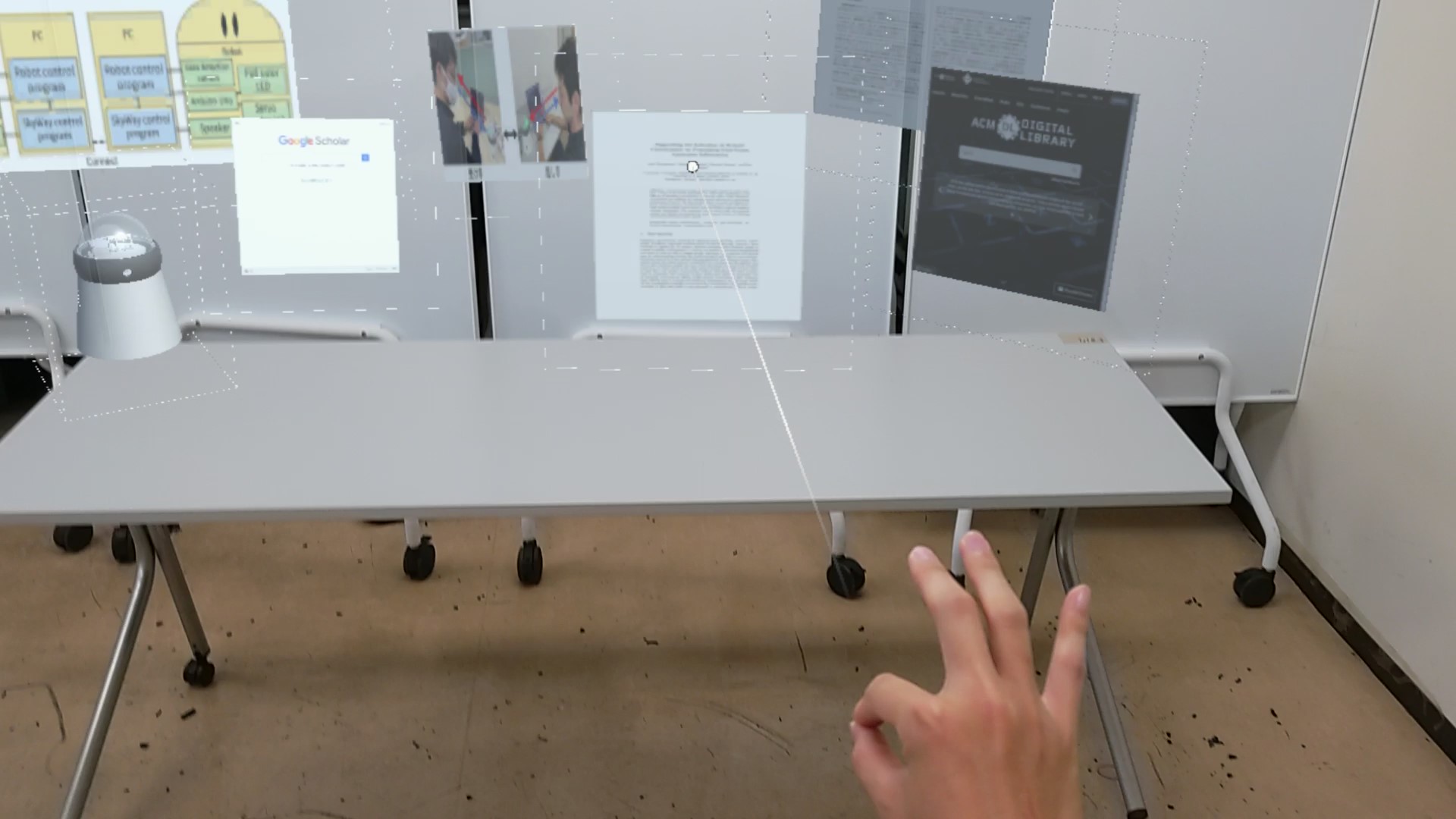} 
\end{minipage}
\begin{minipage}[b]{0.49\linewidth}
\includegraphics[keepaspectratio,scale=0.0605]{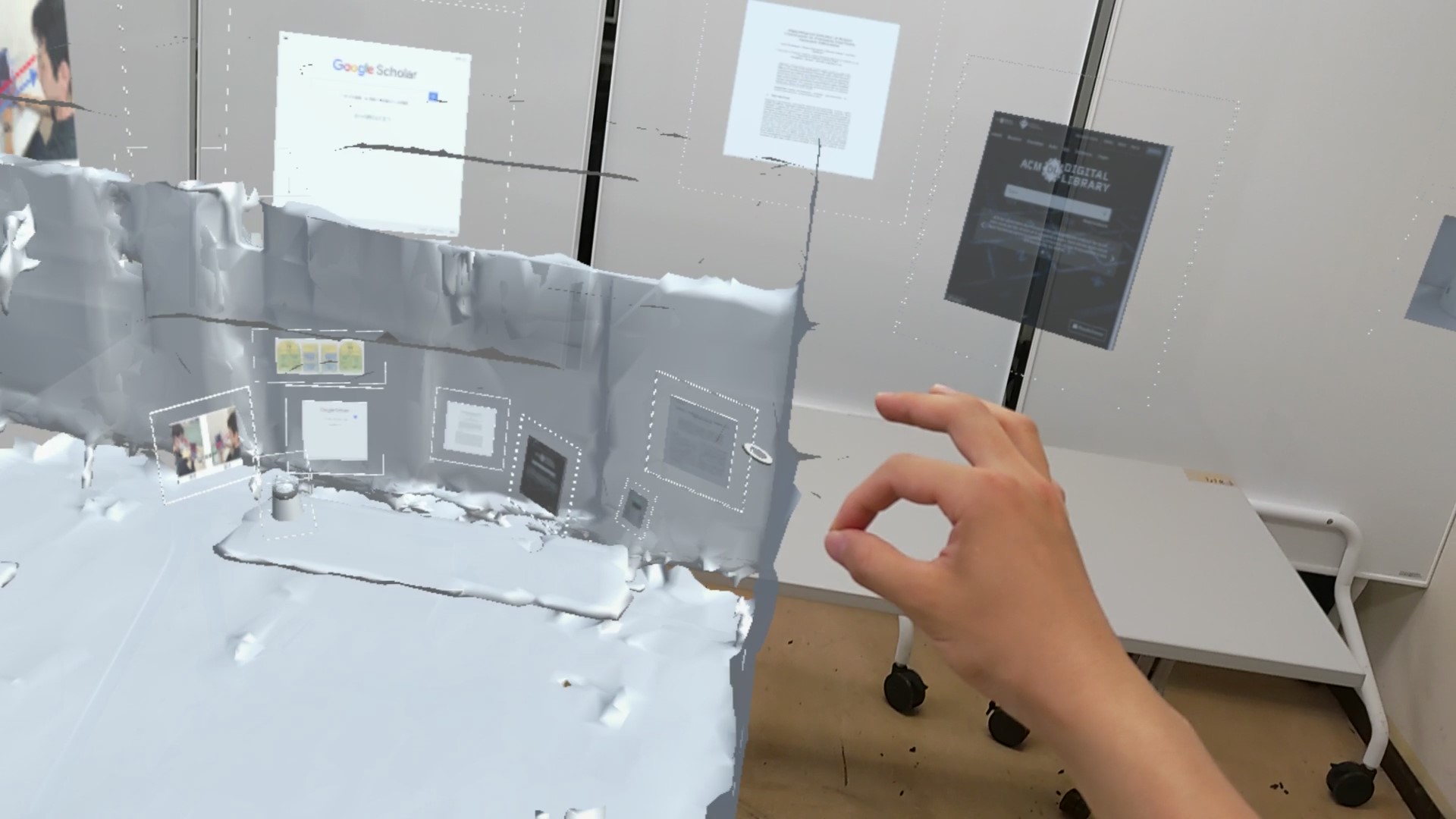} 
\end{minipage}
\caption{\label{condition}Manipulation in \textit{Hand-Ray} (left) and \textit{Miniature} (right).}
\end{center}
\end{figure}

In \textit{Hand-Ray}, participants used a ray emitted from their hand to select virtual objects. We compared this with \textit{Miniature} because this is a general method in AR, and this is the default manipulation method in HoloLens2.
In \textit{Miniature}, we asked the participants to use only the miniature manipulation.
In miniature manipulation, participants can change the size and position of the miniature room and miniature objects.
Figure \ref{condition} shows the manipulation of each condition.

\subsection{Study design}

\begin{figure}[H]

\begin{center}
\includegraphics[width=0.3\textwidth]{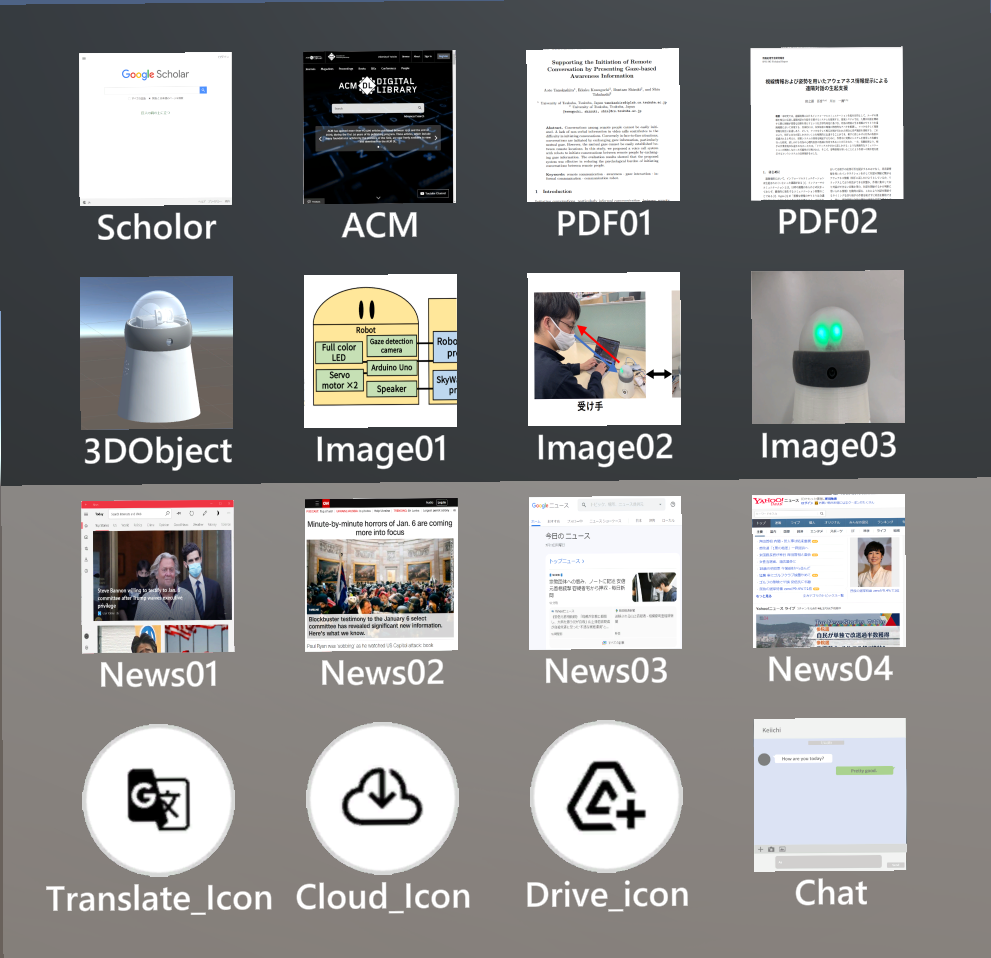} 
\caption{\label{16objects}Sixteen objects used in our task. The top 8 are for the Productivity scenario, and the bottom 8 are for the Leisure scenario.}
\end{center}
\end{figure}

We used a task in which the participants place virtual objects satisfying the scenarios we provided.
Furthermore, we used two scenarios based on previous research\cite{SemanticAdapt}.
The first scenario is a Productivity scenario in which a user is surveying papers for a particular topic.
The second scenario is a Leisure scenario in which a user gathers news information while chatting with a peer.
For each scenario, we used eight virtual objects (Fig.\ref{16objects}).
We used web browsers, PDF readers, 3D objects, and image windows for the Productivity scenario.
We used news readers, 3D icons, and a chat window for the Leisure scenario.
In our task, we placed two desks in a room, and asked the participants to place virtual objects around those desks for each scenario. 
Therefore, in each task, participants placed eight objects on each desk, and a total of 16 objects were placed.


\subsection{Participants}
We recruited 12 undergraduate/graduate students of computer science from our university (10 male, 2 female: $M=23.17$, $SD=0.83$).
Participants answered a questionnaire about how much were they familiar with VR/AR on a Likert scale from 1 (not at all familiar) to 5 (very familiar), and the mean score was 2.67 ($SD = 1.37$).

\subsection{Procedure}
First, participants performed a training session to familiarize with UI manipulation and both \textit{Hand-Ray} manipulation and \textit{Miniature} manipulation.
In the training session, participants selected a sphere from the UI and practiced moving and scaling manipulation on the sphere until they were satisfied.
Second, we conducted the tasks.
We asked participants to place the virtual objects until they had completed their placement that satisfied the two scenarios.

\begin{figure}[H]
\begin{center}
\begin{minipage}[b]{0.45\linewidth}
\includegraphics[keepaspectratio,scale=0.0265]{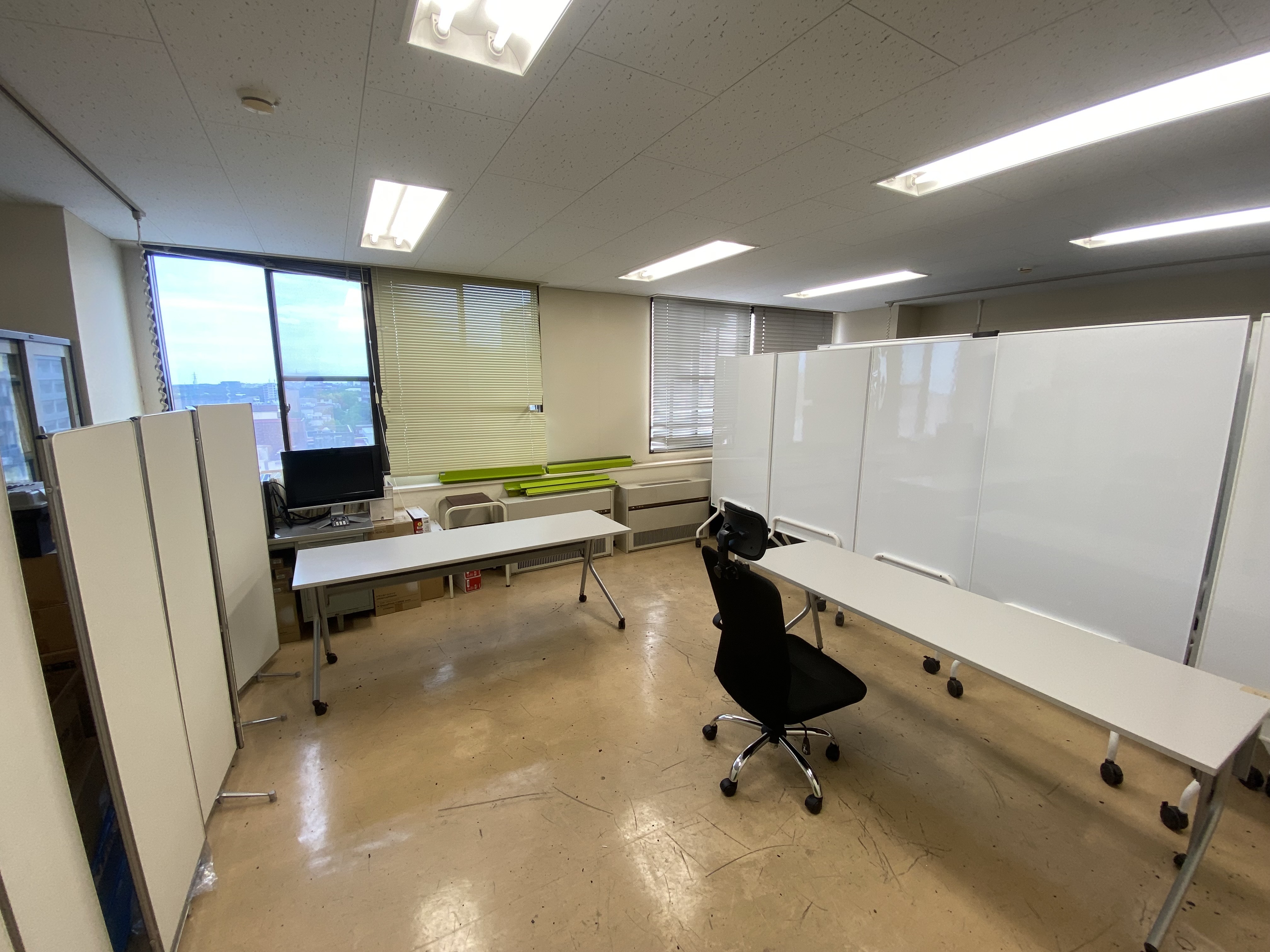} 
\end{minipage}
\begin{minipage}[b]{0.45\linewidth}
\includegraphics[keepaspectratio,scale=0.0265]{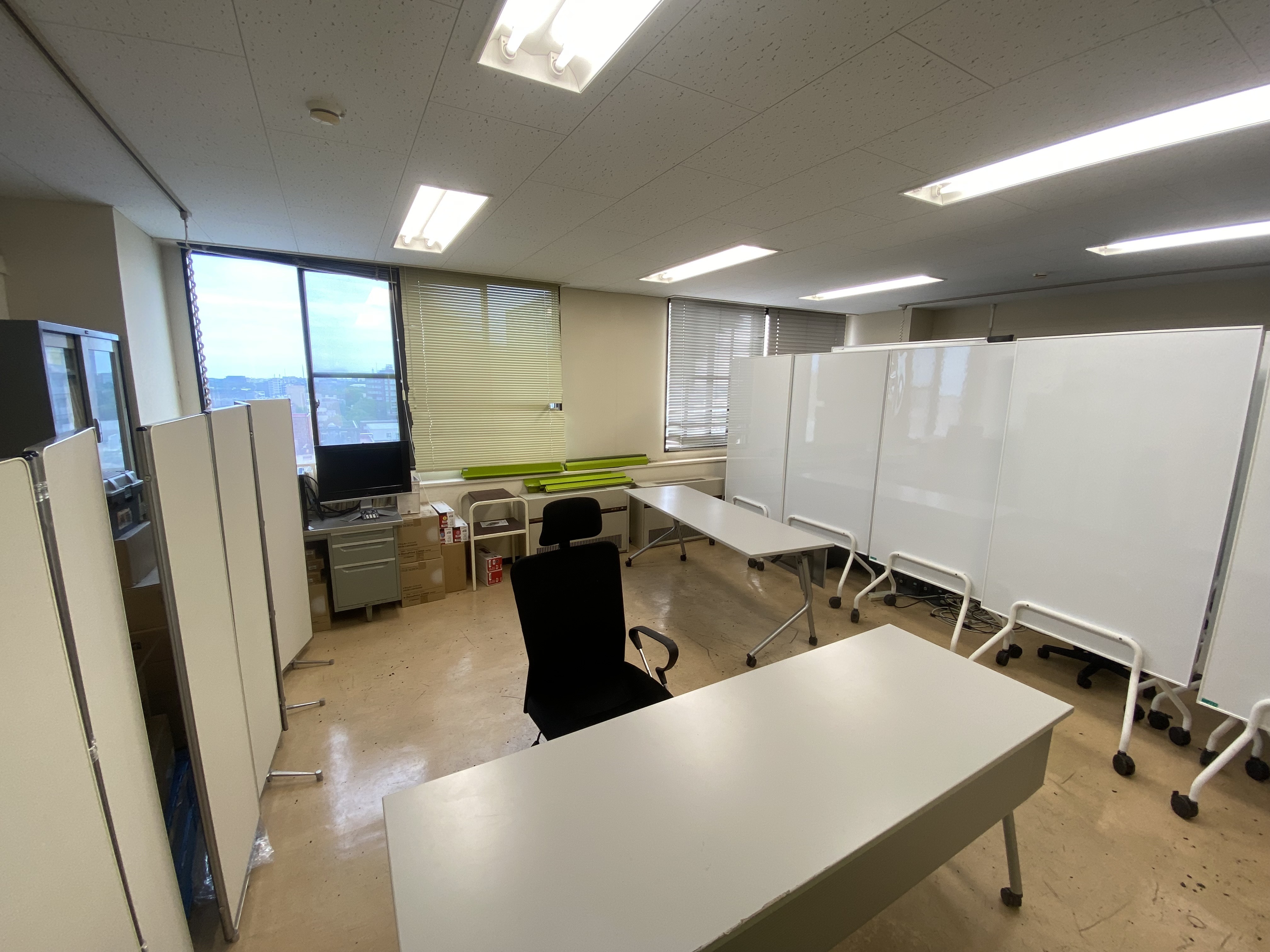} 
\end{minipage}
\caption{\label{miniature_room_manipulation}Desk arrangements, A1 (left) and A2 (right).}
\end{center}
\end{figure}

In our task, we did not ask the participants to make a particular AR layout but asked them to make an AR layout as they want.
We consider that if we ask them to make the layout as fast as they can, they might make the layout roughly.
Therefore, we ask them to keep manipulating the virtual objects until they are satisfied with the layout.
In one task, we used two scenarios, and the participants placed 16 objects.
We used a within-subjects design, and the participants conducted the same task using the same two scenarios and the same 16 objects under two conditions.
We assumed that if the desk arrangement for the two tasks were the same, participants would not consider the arrangement again in the second task.
Therefore, for each task, we used two different desk arrangements, A1 and A2 (Fig.\ref{miniature_room_manipulation}).
To counterbalance, 3 participants conducted the tasks in the order of A1/\textit{Hand-Ray} and A2/\textit{Miniature},
3 participants in the order of A1/\textit{Miniature} and A2/\textit{Hand-Ray},
3 participants in the order of A2/\textit{Hand-Ray} and A1/\textit{Miniature}, and
3 participants in the order of A2/\textit{Miniature} and A1/\textit{Hand-Ray}.
Finally, after the task, we conducted questionnaires.
\subsection{Measurement}
As we considered that there is a difference in manipulation time and workload between the two conditions, we obtained the following data.

\subsubsection{Objective measures}
We obtained the time from the start of the task to the completion of the placement of the 16 objects, which we defined as the total time of the manipulation.
In addition, we considered that object placement can be divided into three parts: moving, scaling, and confirmation of placement. 
Therefore, we obtained the time of these three parts.  
Because the movement manipulation was performed using one hand, we measured the time spent holding virtual objects with one hand and used the total time as the moving manipulation time.
Similarly, because scaling manipulation was performed using two hands, we measured the time spent holding virtual objects with two hands and used the total time as the scaling manipulation time.
Furthermore, we used the rest of the time as the confirmation of placement time.

\subsubsection{Subjective measures}
We conducted the NASA-TLX workload questionnaire\cite{nasatlx} to measure the workload, and the System Usability Scale (SUS)\cite{sus} to measure the usability of the condition.
To measure the satisfaction of the AR layouts, we conducted an additional questionnaire based on previous research~\cite{DistanciAR}.
The additional item was "it was easy to place objects at the desired location " (Q1) and "it was easy to create my intended AR layout" (Q2), and we used a 7-point Likert scale (1: strongly disagree, 7: strongly agree).
In addition, we conducted an open-ended questionnaire.

\section{Results}

We conducted a statistical analysis of the results obtained from the experiment.
In this experiment, two factors might have affected the results: conditions (\textit{HandRay} and \textit{Miniature}) and desk arrangements (A1 and A2).
Although, as statistical analysis showed no difference between desk arrangements A1 and A2 in manipulation time, NASA-TLX, and SUS, we considered that the desk arrangements did not affect the manipulation.
Therefore, we conducted a one-factor analysis considering only the conditions: \textit{Hand-Ray} and \textit{Miniature}.
We performed a t-test for those in which the data followed a normal distribution.
In contrast, we performed a Wilcoxon signed rank test for those of which the data did not follow a normal distribution.
We set the significance level at 5\%.
In the graph, we marked * when $p<=0.05$, and ** when $p<=0.01$.

\subsection{Manipulation time}
\begin{figure*}[h]
\begin{center}
    \begin{minipage}[h]{0.485\linewidth}
    \begin{center}
    \includegraphics[width=0.95\textwidth]{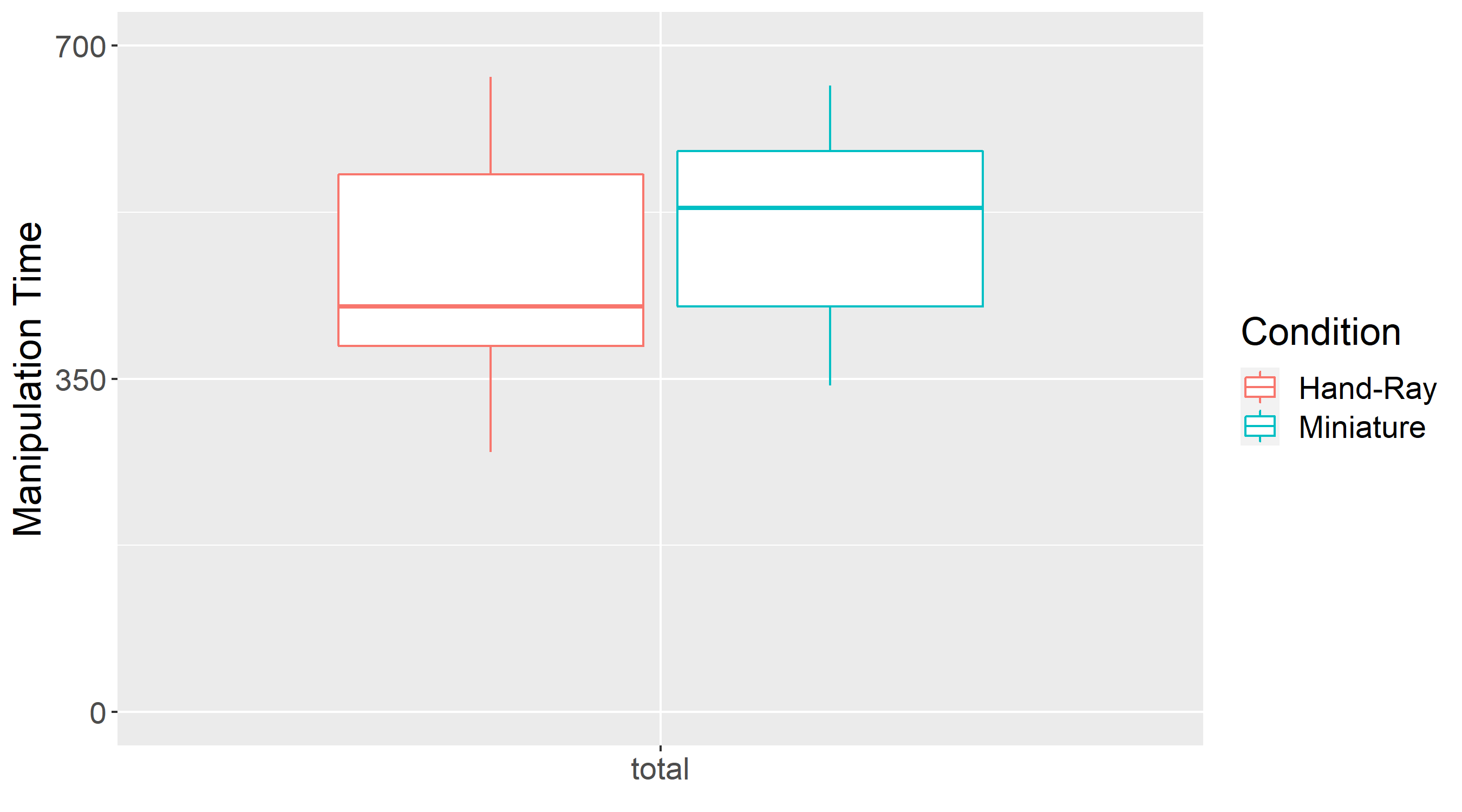} 
    \subcaption{\label{total_time}Total manipulation time.}
    \end{center}
    \end{minipage}  
    \begin{minipage}[h]{0.475\linewidth}
    \begin{center}
    \includegraphics[width=0.95\textwidth]{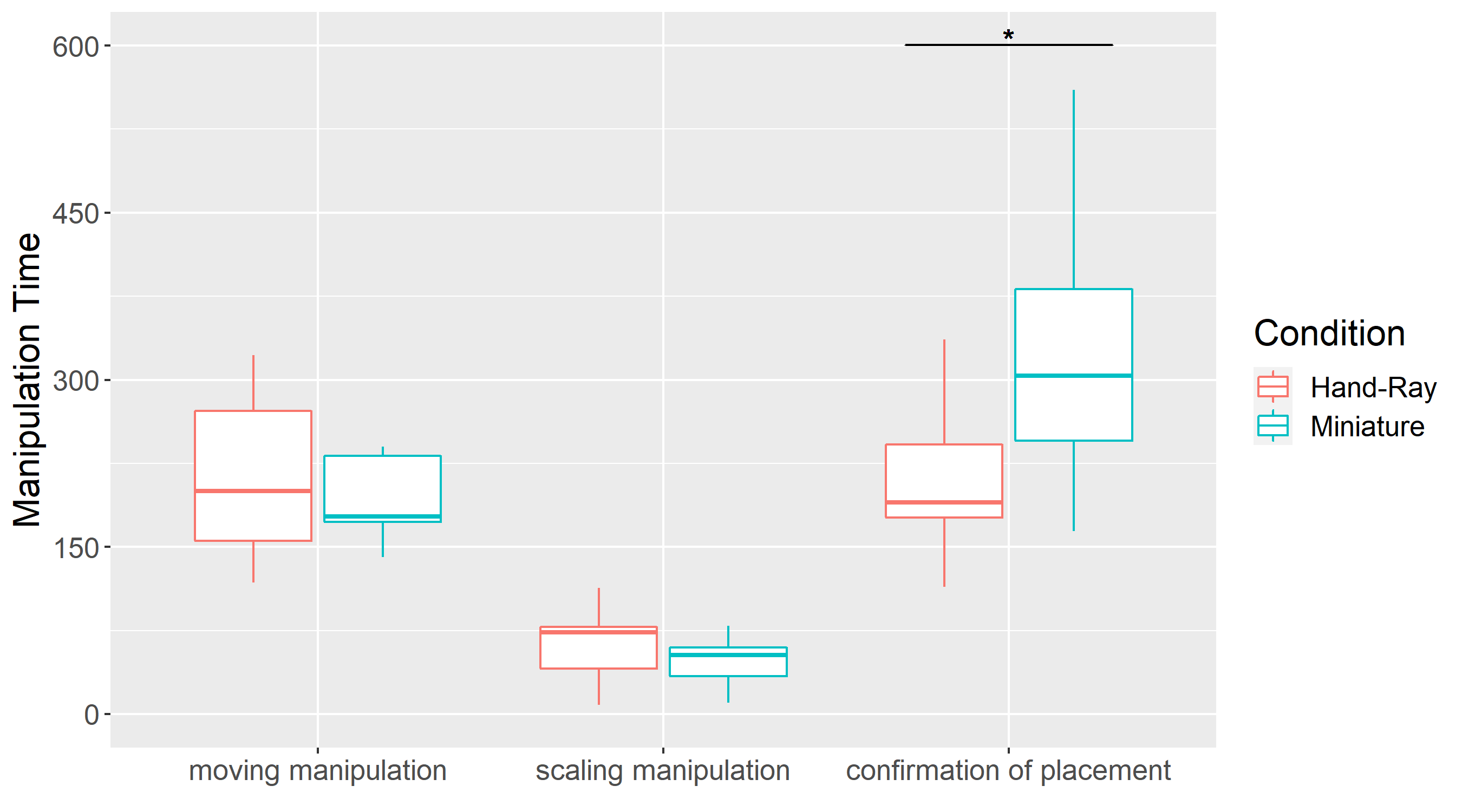} 
    \subcaption{\label{each_time}Individual manipulation time.}
    \end{center}
    \end{minipage}  
\end{center}
\caption{\label{mani}Manipulation time for each condition.}
\end{figure*}

First, we analyzed the results of the total manipulation time.
The results are shown in Fig.\ref{total_time}.
The mean of the total manipulation time was 560 s for \textit{Hand-Ray} ($SD=347$), and 593 s ($SD=210$) for \textit{Miniature}.
A Wilcoxon sign test showed no significant difference ($V=22$,  $p=0.204$).

Next, we analyzed the results of each manipulation, moving manipulation, scaling manipulation, and confirmation of placement.
The results are shown in Fig.\ref{each_time}.
The mean of the moving manipulation time was 225 s ($SD=97.5$) for \textit{Hand-Ray}, and 221 s ($SD=86.8$) for \textit{Miniature}.
A Wilcoxon sign test showed no significant difference ($V=37$, $p=0.910$).
The mean of the scaling manipulation time was 60.8 s ($SD=30.0$) for \textit{Hand-Ray}, and 52.6 s ($SD=28.3$) for \textit{Miniature}.
A paired t-test showed no significant difference ($t(11)=1.05$, $ p=0.3142$).
The mean of the confirmation of placement time was 275 s ($SD=244$) for \textit{Hand-Ray}, and 320 s ($SD=123$) for \textit{Miniature}.
A Wilcoxon sign test showed a significant difference ($V=13$, $p=0.0425 < 0.05$).

\begin{figure*}[h]
\begin{center}
    \begin{minipage}[h]{0.54\linewidth}
    \begin{center}
    \includegraphics[width=0.95\textwidth]{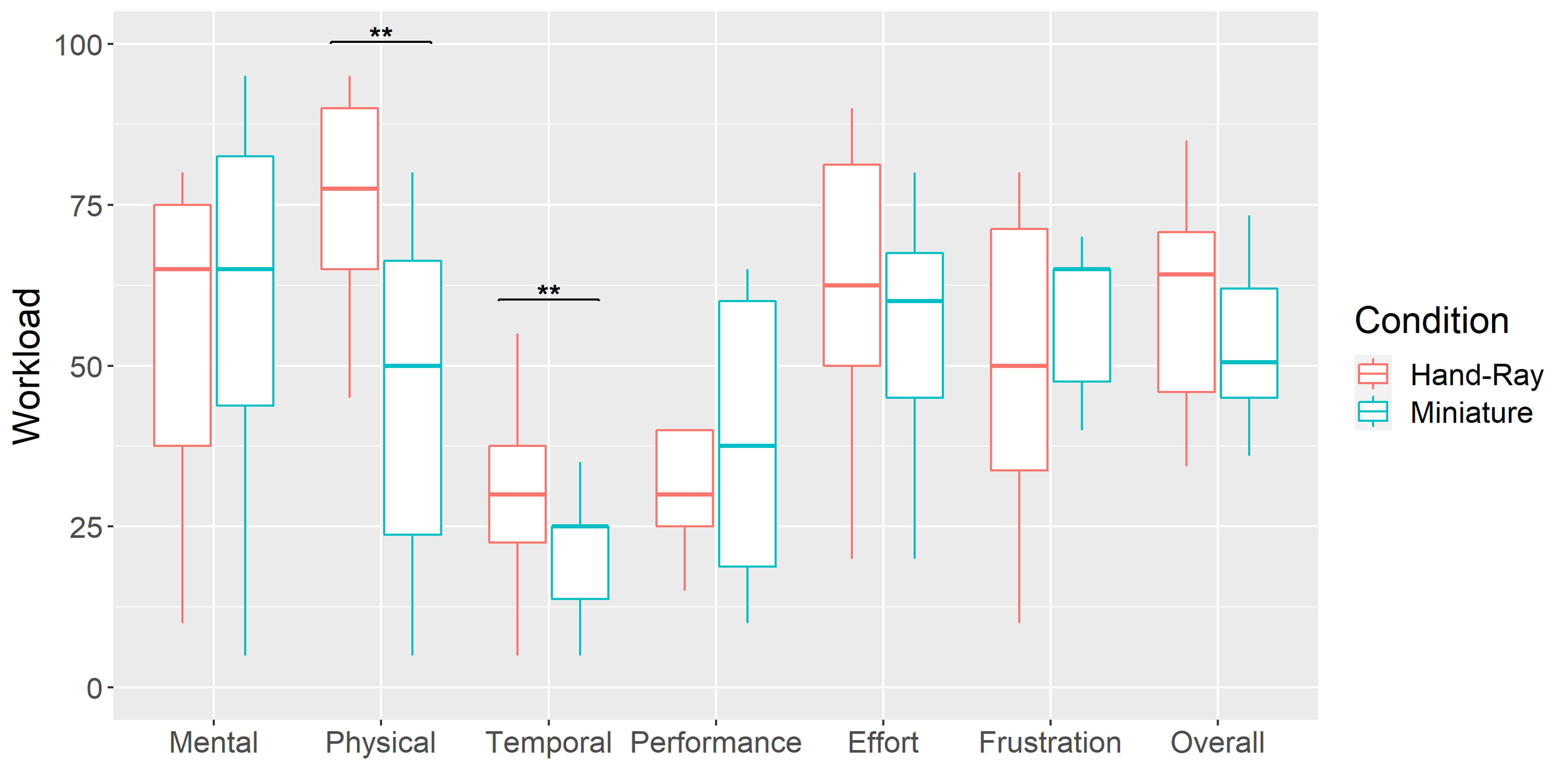} 
    \subcaption{\label{fig:nasa-tlx}NASA-TLX results.}
    \end{center}
    \end{minipage}  
    \begin{minipage}[h]{0.42\linewidth}
    \begin{center}
    \includegraphics[width=0.95\textwidth]{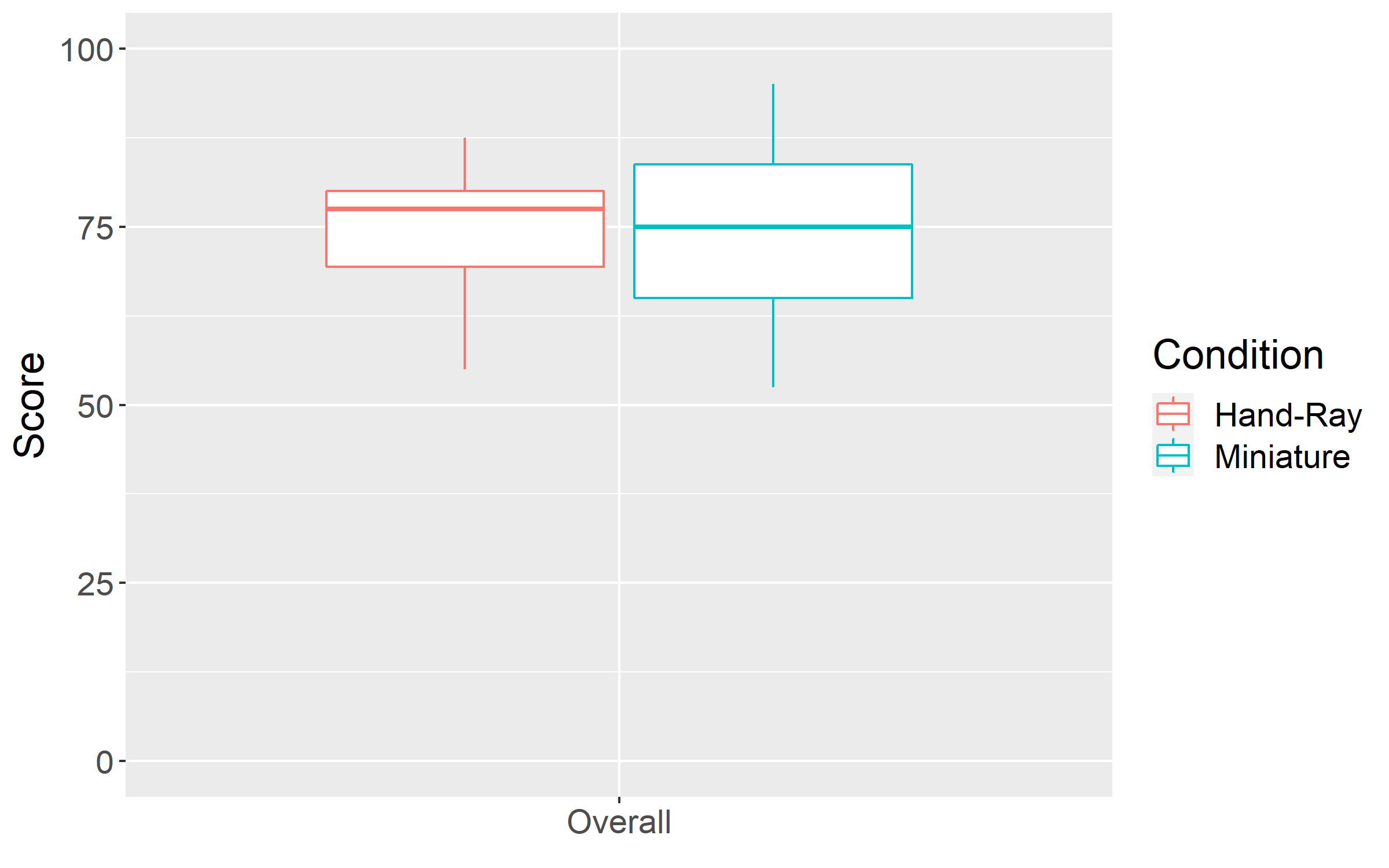} 
    \subcaption{\label{sus_score}SUS results.}
    \end{center}
    \end{minipage}  
\end{center}
\caption{\label{mani}Subjective measurement results for each condition: (a) NASA-TLX, (b) SUS.}
\end{figure*}
\newpage
\subsection{Workload}
The results of the NASA-TLX workload are shown in Fig.\ref{fig:nasa-tlx}.
In NASA-TLX, a higher score indicates a higher workload.
First, we analyzed the overall score calculated from each category.
The mean of the overall score was 60.2 ($SD=16.8$) for \textit{Hand-Ray}, and 49.8 ($SD=18.2$) for \textit{Miniature}.
A paired t-test showed no significant difference ($t(11)=2.01$,  $p=0.0691$).

Next, we analyzed each category.
For Mental demand, the mean score was 55.4 ($SD=25.3$) for \textit{Hand-Ray} and 60.8 ($SD=28.4$) for \textit{Miniature}.
A Wilcoxon sign test showed no significant difference ($V=27.5$, $ p=0.386$).
For Physical demand, the mean score was 75.8 ($SD=17.2$) for \textit{Hand-Ray} and 45.5 ($SD=25.9$) for \textit{Miniature}.
A paired t-test showed a significant difference ($t(11)=3.50$, $p=0.00513 < 0.01$).
For Temporal demand, the mean score was 30.4 ($SD=15.4$) for \textit{Hand-Ray}, and 22.1 ($SD=13.9$) for \textit{Miniature}.
A paired t-test showed a significant difference ($t(11)=3.25$, $p=0.00771 < 0.01$).
For Performance, the mean score was 37.9 ($SD=22.1$) for \textit{Hand-Ray} and 37.1 ($SD=20.9$) for \textit{Miniature}.
A Wilcoxon sign test showed no significant difference ($V=29$, $p=0.919$).
For Effort, the mean score was 62.1 ($SD=23.2$) for \textit{Hand-Ray} and 53.4 ($SD=23.0$) for \textit{Miniature}.
A paired t-test showed no significant difference ($t(11)=0.969$,  $p=0.354$).
For Frustration level, the mean score was 50.4 ($SD=22.2$) for \textit{Hand-Ray} and 52.9 ($SD=21.0$) for \textit{Miniature}.
A Wilcoxon sign test showed no significant difference ($V=31$, $p=0.89$).

\subsection{Usability}
The results of the overall SUS score are shown in Fig.\ref{sus_score}.
In SUS, a higher score indicates higher usability.
For SUS score, the mean score was 73.3 ($SD=13.2$) for \textit{Hand-Ray} and 73.8 ($SD=14.4$) for \textit{Miniature}.
A Wilcoxon sign test showed no significant difference ($V=31$, $p=0.89$).

\subsection{Additional questionnaire}
\begin{figure}[H]
\begin{center}
\includegraphics[width=0.5\textwidth]{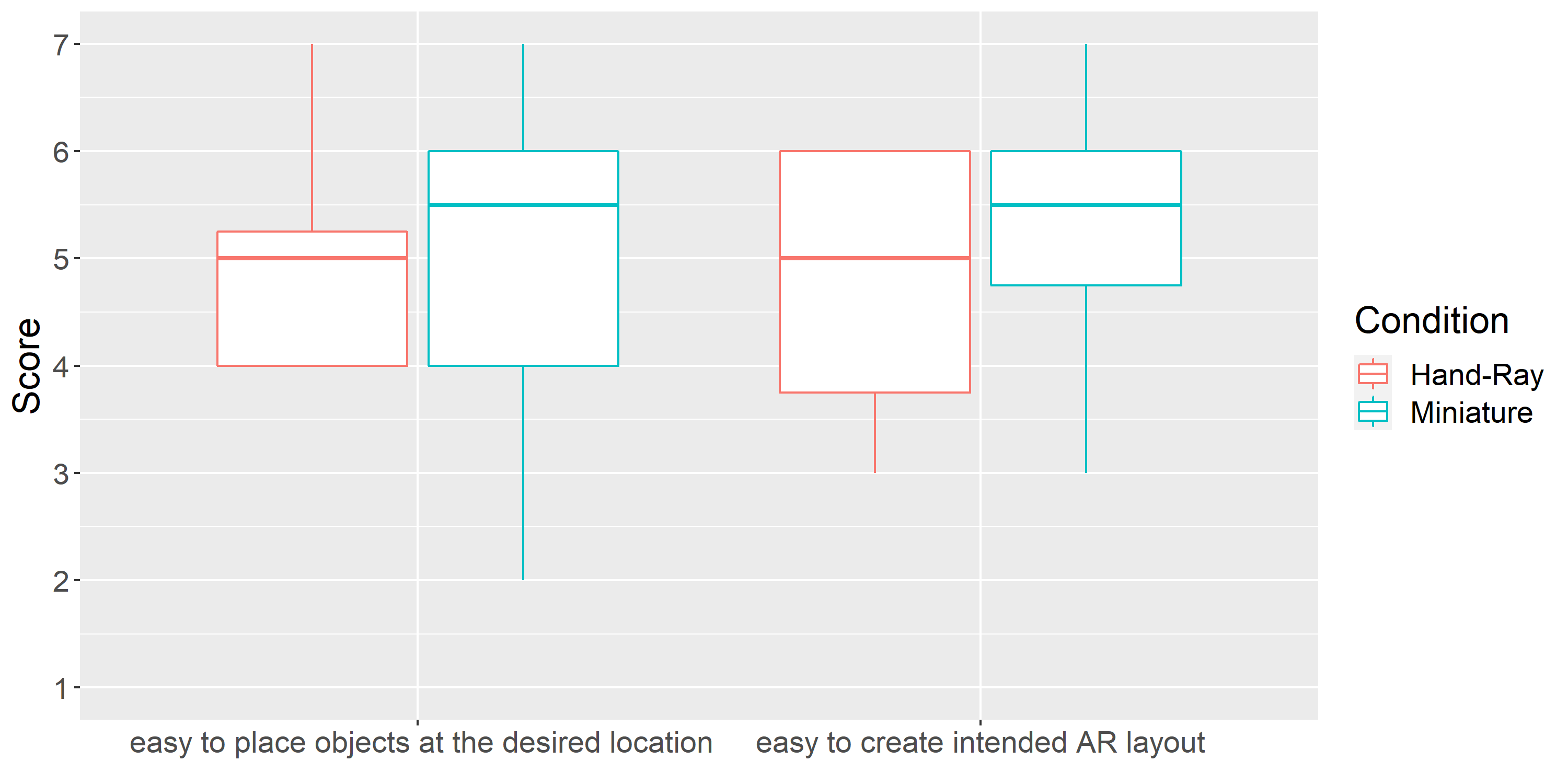} 
\caption{\label{additional}Additional questionnaire results for each condition.}
\end{center}
\end{figure}

The results of the additional questionnaire are shown in Fig.\ref{additional}.
For "it was easy to place objects at the desired location" (Q1), the mean score was 4.58 ($SD=1.51$) for \textit{Hand-Ray} and 5.08 ($SD=1.56$) for \textit{Miniature}.
A paired t-test showed no significant difference ($t(11)=-0.920$, $p=0.377$).
For "it was easy to create my intended AR layout" (Q2), the mean score was 4.75 ($SD=1.22$) for \textit{Hand-Ray} and 5.17 ($SD=1.27$) for \textit{Miniature}.
A Wilcoxon sign test showed no significant difference ($V=15$, $p=0.401$).
\newpage

\section{Discussion}
\subsection{Physical workload}
NASA-TLX results showed that our system is less physically demanding than hand-ray manipulation.
In addition, the open-ended questionnaire provided the following comments regarding physical workload.
"In \textit{Miniature}, my arms was not fatigued compared to \textit{Hand-Ray}. I think this is because the miniature can be placed in the desired position, and the manipulation can be performed where the arms would not get tired" (P3).
"In \textit{Miniature}, I did not have to raise my hands, so my arms were less tired" (P5).
"In \textit{Miniature}, I could move objects with ease because I did not have to move my hands forward in a big motion" (P10).
As users can change the position of the miniature, they can manipulate objects without raising their arms or stretching them forward.
This might be one of the factors that our system reduced the physical workload. 

In addition, we received the following comments.
"In \textit{Miniature}, I did not have to move my hands a lot, so I could manipulate objects without getting tired" (P6).
"In \textit{Miniature}, all manipulations can be performed with small movements of my hand, so my arm was not fatigued" (P12).
Our system allows users moving and scaling manipulations to be performed with small hand-movements.
This might be another factor that our system reduced the physical workload.

\subsection{Time pressure}
NASA-TLX results showed that our system is less temporally demanding than hand-ray manipulation. 
This result indicates that our system puts less time pressure on the participants.
This result is interesting because the confirmation of placement time of our system was significantly longer than hand-ray manipulation (Fig.\ref{each_time}), and the total manipulation time was not significant but also longer than hand-ray manipulation (Fig.\ref{total_time}).
In the open-ended questionnaire, we obtained the following comments.
"In \textit{Miniature}, it was easy to understand the size and position of virtual objects due to the bird's eye view of the entire room" (P12).
"In \textit{Miniature}, as I had a bird's eye view of the entire project, it was easy to place the whole thing in balance" (P6).
"In \textit{Miniature}, I did not have to shake my head from side to side to find the placement" (P1).
These comments suggest that, in \textit{Miniature}, it might be easy to grasp the entire layout.
Therefore, in \textit{Miniature}, participants may have been able to estimate the time for how soon they could complete the task, and this might decrease the temporal demand despite the total manipulation time and the confirmation of placement time increased.

\subsection{Confirmation of placement}
\textit{Hand-Ray} took less time in the confirmation of placement time than \textit{Miniature}.
In addition, the open-ended questionnaire provided the following comments regarding the confirmation of placement.
"Even if I think that this must be the right place in the miniature when I actually look at it in the real environment, it is not quite right" (P11).
"In \textit{Miniature}, after placing the object in the miniature, I had to turn to the real-scale object to check if it was placed in the correct position and then look at the miniature to adjust it, then check it, and so on" (P4).
"Sometimes the orientation of the miniature objects was different from what I expected when I actually look them in the real environment, and it was surprisingly tedious to correct them" (P5).
These comments suggest that, in our system, users must check the virtual objects alternately in the real environment and in the miniature.
We considered that this might increase the confirmation of placement time.

\subsection{Moving/Scaling manipulation}
Our system requires less hand movement than hand-ray manipulation to perform the moving and scaling manipulation.
However, there was no difference between \textit{Miniature} and \textit{Hand-Ray} in these manipulation time.
In the open-ended questionnaire, the following responses were obtained regarding these manipulations.
"When moving objects large distances, \textit{Miniature} was easier because the amount of hand movement was smaller. However, when moving objects in detail, it was easier to use \textit{Hand-Ray}" (P5).
"In \textit{Miniature}, it was easy to move objects to a certain position, but it was difficult to arrange them neatly from there" (P4).
These comments suggest that large moves and scaling may be better with \textit{Miniature}, and fine moves and scaling may be better with \textit{Hand-Ray}.

\section{Limitations and Future Work}
\subsection{Manipulation mode}
In our system, users switch the manipulation mode between miniature room manipulation and miniature objects manipulation using a virtual button in a hand menu.
In addition, users can understand the current mode by checking whether the bounding box is attached to the miniature room or not.
On these manipulations, we obtained the following comments in the open-ended questionnaire.
"I feel that if it were easier to change the manipulation mode of the miniature, it would further enhance the feeling of the manipulation" (P4).
"Sometimes I was confused about which mode I was in now" (P7).
From these comments, by making it easier to change the manipulation mode and check the current mode, we can improve the usability of our system.
For example, we can display the switch button in or near the miniature to ensure that the users can switch the mode easily.
In addition, we can change the color of the miniature depending on the current mode to ensure that the users can check the current mode easily.

\subsection{Situation where the miniature is difficult to see}
We obtained the following comments in the open-ended questionnaire.
"Sometimes the miniature was blocked by virtual objects, and I could not see the miniature" (P15).
"There were times when the miniature and the virtual objects in the real environment overlapped, and I could not figure out what was going on" (P7).
These comments indicate a problem in our system that when the users place a virtual object in the real environment near the miniature, they have difficulty seeing the miniature.
Therefore, in our future work, we will make virtual objects near the miniature translucent or transparent to solve this problem.

\subsection{Combination of manipulations}
In this experiment, we did not consider manipulation using both the miniature and hand-ray techniques.
However, because these manipulations do not interfere, they can be combined.
Using the experiment results, we can see how to combine them.
In moving manipulation, using miniature may be better for large moves, and using hand-ray may be better for fine moves.
Therefore, it might be appropriate to first move roughly using miniature and then move in detail using hand-ray.
In addition, because our system requires considerable time to alternate between looking at the miniature and real environments, confirmation of placement time requires more time with our system than hand-ray manipulation.
Although, this time can be reduced if detailed manipulation is performed by hand-ray manipulation.

The following is an example of how both manipulations can be combined. 
\begin{enumerate}
    \item Select multiple virtual objects from the hand menu and display objects you want to use.
    \item Roughly manipulate objects by manipulating miniature objects.
    \item Perform detailed manipulation using hand-ray.
    \item When changing the working place, roughly manipulate objects by manipulating miniature objects.
    \item Perform detailed manipulation using hand-ray.
    \item Repeat this procedure.
\end{enumerate}

In future work, we will create a system that combines hand-ray manipulation and miniature manipulation in this manner and conduct an evaluation experiment.

\subsection{Study design}
In the task of our evaluation, we asked the participants to make AR layouts for the two scenarios.
One is a Productivity scenario, in which participants make a layout for surveying documents, and the other is a Leisure scenario, in which participants make a layout for relaxing with others.
The task is just until the fixing of the layout, and the actual use of the layout (i.e. conducting a survey, relaxing with others) is outside the scope of this experiment.
How users behave in a fixed AR layout is also an interesting topic, so in the future work, it might be important to evaluate actual usage in these scenarios.
\section{Conclusion}
In this study, we presented a system to reduce the burden on users in AR layout.
In our system, we created a miniature using a mesh of a room acquired by the depth sensors attached to the AR device and made the WIM technique available in AR.
Furthermore, we compared our system with hand-ray manipulation to compare manipulation time and workload.
The results showed that our system significantly reduced workload in physical and temporal demand.
The open-ended questionnaire suggested that the position of the miniature and amount of hand movement might affect the physical demands.
In addition, the ease of grasping the entire layout might affect temporal demand.
However, the results showed no significant difference in manipulation time.
Conversely, the confirmation of placement time for our system was significantly longer than the hand-ray manipulation.
A possible factor is that, in our system, users need to alternately check the real and miniature environments during placement.
As a limitation, we did not consider a system combining the miniature and hand-ray techniques in this study.
Therefore, in future work, we will develop such a system considering the advantages and disadvantages of our system revealed by the experiment.

\bibliographystyle{eg-alpha-doi}
\bibliography{egbibsample}

\end{CJK}
\end{document}